# Interferometry of a Single Nanoparticle Using the Gouy Phase of a Focused Laser Beam

Jaesuk Hwang [a,b,+], W. E. Moerner [a]

[a] Department of Chemistry, Stanford University, Stanford, CA 94305 USA

[b] Department of Applied Physics, Stanford University, Stanford, CA 94305 USA

Abstract: We provide a quantitative explanation of the mechanism of the far-field intensity modulation induced by a nanoparticle in a focused Gaussian laser beam, as was demonstrated in several recent direct detection studies. Most approaches take advantage of interference between the incident light and the scattered light from a nanoparticle to facilitate a linear dependence of the signal on the nanoparticle volume. The phase relation between the incoming field and the scattered field by the nanoparticle is elucidated by the concept of Gouy phase. This phase relation is used to analyze the far-field signal-to-noise ratio as a function of exact nanoparticle position with respect to the beam focus. The calculation suggests that a purely dispersive nanoparticle should be displaced from the Gaussian beam focus to generate a far-field intensity change.

[+]Current Address: Laboratory for Physical Chemistry, ETH Zuerich, Switzerland

Ultrasensitive optical spectroscopy of condensed phases in fields ranging from physical chemistry[1,2] to single-molecule biophysics[3,4] often make use of emission from highly fluorescent guest molecules such as organic dyes[5,6], or autofluorescent proteins[7]. Although for a number of reasons fluorescence methods are currently widely used in single-molecule studies, sensitive detection of fluorescence in condensed matter also presents several experimental issues, most notably the need to count single photons while rigorously excluding counts from impurity fluorescence or Raman scattering from trillions of host molecules. These issues are even more challenging in non-artificial experimental environments such as inside cells or turbid hosts. An alternative detection method is direct measurement of absorption, in which the absorption events are not detected by recording subsequent fluorescence but by the change in the power or phase of a laser beam probing the sample. In this case, any spurious emission from the sample or substrate is not critical, and all the photons in the entire laser beam probing the sample can be used to sense the signal of interest. Frequency-modulation (FM) spectroscopy[8] was the first method to allow detection of the absorption of single molecules in condensed matter[9], and low temperatures were required to provided narrow absorption lines. At room temperature, a new type of absorption spectroscopy based on Sagnac interferometry[10] demonstrated detection sensitivity down to a few absorbing molecules. In both of these measurements, interference between the forward signal portion of the beam and the local oscillator beam played a key role in linearizing the response from the small signal to overcome electronic noise of the detection system.

Recently, various direct detection schemes were introduced that take advantage of interference between the forward scattered light from a nanoparticle and the incident light. Via common-path interference, a better signal-to-noise ratio than the conventional scattering measurement[11-16] was achieved mainly because the signal was proportional to the volume of the nanoparticle rather than the volume squared. Figure 1 illustrates the reflection and transmission geometries of such experiments. This type of single beam interferometry of small particles was first suggested by Batchelder and Taubenblatt[17]. Although the basics of the method relies on interference phenomena, the detailed explanation can be subtle due to non-trivial phase behavior near the waist of a focused laser beam. Traditionally, the measurements of light scattered from a small particle have been performed at an angle that avoids the collection of the original excitation light[18, 19]. For this reason, the theoretical approaches paid little attention to the phase because most of the experiments measured only the intensity of the scattering and few experiments were performed to measure the amplitude and phase of the scattering in all directions $\theta$ and $\phi$. However, for interference measurements, a definite phase relation between the incident and the scattered beam is a key requirement to fully interpret the signal. Therefore, it is the goal of this paper to theoretically study the phase aspects of the interference between a focused Gaussian laser beam and the scattered beam from a nanoparticle placed therein. It will be shown that the exact phase relation between the two is mainly governed by the geometrical extent of each beam and also that the phenomena of absorption and phase shift are merely the manifestations of the phase relation in coherent addition of scattered light and incident light. In elucidating the physical origins of these phenomena, the concept of Gouy phase[20] plays a key role. This approach will facilitate the prediction of the far-field signal-to-noise ratio behavior induced by a nanoparticle in a focused Gaussian beam as a function of the exact nanoparticle position with respect to the waist.

We generalize the argument by including the phaseshift $\phi$ of the probe beam caused by the nanoparticle as well as the absorption, characterized by the absorption cross section divided by the spot size of the laser beam $\sigma/A$. The optical theorem[18, 19] dictates that the extinction is composed equally of scattering and absorption, and also that the apparent absorption, which is the attenuation of the excitation beam in the far-field, is only the result of the destructive interference between the scattered beam and the forward propagating excitation beam. So by adopting a general cross section $\sigma = \sigma_{abs} = \sigma_{sca}$, the formalism will be developed emphasizing on the interference of the scattering with the incident beam. From here on $\sigma/A$ will quantify the strength of scattering, rather than absorption. Then it will be shown later that the scattering indeed leads to absorption in the far-field. In the small-signal limit, the on-axis electric field after interaction with the object with phase shift $\phi$ can be written as

$$\begin{aligned} E(z) &= E_0 \exp(ikz)(1+\frac{\sigma}{A})\exp(i\phi) \\ &\cong E_0 \exp(ikz)(1+\frac{\sigma}{A})(1+i\phi) \\ &\cong E_0 \exp(ikz) + \frac{\sigma}{A} E_0 \exp(ikz) + i\phi E_0 \exp(ikz) \end{aligned} \quad (1)$$

Here, the time-varying factor $e^{-i\omega t}$ is assumed. We pay careful attention to the phase development of each portion of the beam as the geometry deviates from the plane wave, such as near the focus (waist). Therefore, rewriting Equation (1) considering the different phase factors that develop on each portion of the beam:

$$E_{out} \cong E_0 \exp(ikz + i\phi_{G,inc}(z)) + \frac{\sigma}{A} E_0 \exp(ikz + i\phi_{sc}(z)) + i\phi E_0 \exp(ikz + i\phi_{sc}(z))$$

$$\Delta\phi_G(z) = \phi_{sc}(z) - \phi_{G,inc}(z)$$

(2)

where $\phi_{G,inc}(z)$ is the Gouy phase of the incident probe beam, $\phi_{sc}(z)$ the phase shift of the scattered beam, all compared to that of the plane wave propagation. These phase factors reflect the fact that the incident beam and the scattered beam will propagate in different spatial modes with respect to each other with increasing $z$. The intensity detected in the far-field is given by

$$E_{out} E_{out}^*(z) = E_0^2 (1 + 2\frac{\sigma}{A} \cos \Delta\phi(z) - 2\phi \sin \Delta\phi(z)) \qquad (3)$$

where $\Delta\phi = \phi_{sc} - \phi_{G,inc}$, which is the phase difference far from the focus. One can see that $\Delta\phi$ is a crucial factor that determines the far-field behavior of the signal. We will from now on evaluate $\phi_{G,inc}$ and $\phi_{sc}$ in more detail.

The Gouy phase shift for the incident beam $\phi_{G,inc}(z)$ from the focus $z = 0$, where the radius of curvature is infinite, to the distance $z$ ($z$ is positive for an advance in the laser propagation direction) on axis is given by $-\tan^{-1}\frac{\lambda z}{\pi\omega_0^2}$ [21], where $\omega_0$ is the beam radius at the focus (waist). The quantity $\pi\omega_0^2/\lambda$ is the usual confocal parameter, equal to half of the Rayleigh range $2z_R$. This phase shift approaches a constant value of $-\pi/2$ at a large distance. The physical origin of the Gouy phase[22] can be explained as follows: Transverse spatial confinement of any shape, through the uncertainty principle, causes a spread in the transverse momenta, which changes the expectation value of the axial propagation constant. The corresponding phase shift is called the Gouy phase. Consider a monochromatic wave of frequency $\omega$ and wave number $k = \frac{\omega}{c}$ propagating along the z direction. For an infinite plane wave, the momentum is $z$ directed and has no transverse components. The spread in transverse momentum is zero and hence, by the uncertainty principle, the spread in transverse position is infinite. A finite beam, however, will have a spread in transverse momentum. The Gouy phase shift is the expectation value of the axial phase shift owing to the transverse momentum spread. This Gouy phase shift is well known as the $-\frac{d\pi}{2}$ axial phase shift that a converging Gaussian beam experiences as it passes through its focus. The dimension $d$ equals 1 for a line focus, and equals 2 for a point focus.

For the scattered beam, $\phi_{sc}(z)$ is $+\frac{\pi}{2}$ both in the far-field and the near-field. This can be understood via the Babinet's principle[18][23]. Let us consider a screen with an aperture much smaller than the wavelength of the excitation. As was discussed in the last paragraph, this small aperture acts as a tight transverse confinement, and the diffraction emerging from the opening will develop $-\frac{\pi}{2}$ phase shift. How quickly this phase develops in space depends on the size of the aperture. For the case of an infinitely small aperture, $-\frac{\pi}{2}$ is immediately reached and the simple factor $-i$ can describe the phase jump. It was recently remarked[22] that this explains the reason for additional $-i$ factor in front of the Kirchoff integral[23]; also the reason that the point source diffraction function (Huygens wavelet) is not exactly a spherical wave but a spherical wave with $-i$ phase shift contrary to the common belief[19]. According to Babinet's principle, if

$E(z) = -iE_{diff} \frac{\exp(ikz)}{kz}$ is the electric field at $z$ when the aperture is present, the field satisfies

$$\overline{E}(z) + E(z) = E_{inc}(z) \qquad (4)$$

where $\overline{E}(z)$ is the electric field at $z$ when the complimentary obstacle is in place, and $E_{inc}(z)$ is the incident field[18]. Then,

$$\overline{E}(z) = E_{inc}(z) - E(z) = \exp(ikz + i\phi_{G,inc}) - (-i)E_{diff}\frac{\exp(ikz)}{kz}$$
$$= \exp(ikz + i\phi_{G,inc}) + iE_{diff}\frac{\exp(ikz)}{kz} \qquad (5)$$

where the first term is only the incident field. This demonstrates that the scattered field from a point obstacle should have $+\frac{\pi}{2}$ phase shift instead of that of a point aperture, $-\frac{\pi}{2}$.

We now evaluate $\Delta\phi = \phi_{sc} - \phi_{G,inc}$ for a general case where the nanoparticle is placed at distance $z$ from the focus on axis. This is illustrated in Figure 2. Starting from the position of the nanoparticle $z$, and measured at the detector, which is placed at sufficiently long distance, the Gouy phase shift of the Gaussian probe beam $\phi_{G,inc}(z)$ is $-\frac{\pi}{2} + \tan^{-1}\frac{\lambda z}{\pi\omega_0^2}$. The scattered beam undergoes a Gouy phase shift of $\phi_{sc}(z) = \frac{\pi}{2}$. Therefore, phase difference $\Delta\phi(z)$ is given by

$$\Delta\phi(z) = \pi - \tan^{-1}\frac{\lambda z}{\pi\omega_0^2}. \qquad (6)$$

Finally the general expression of far-field intensity as a function of the nanoparticle position can be written

$$E_{out}E_{out}^{*}(z) = E_0^2\left(1 + 2\frac{\sigma}{A}\cos\left(\pi - \tan^{-1}\frac{\lambda z}{\pi\omega_0^2}\right) - 2\phi\sin\left(\pi - \tan^{-1}\frac{\lambda z}{\pi\omega_0^2}\right)\right)$$
$$= E_0^2\left(1 - 2\frac{\sigma}{A}\cos\left(\tan^{-1}\frac{\lambda z}{\pi\omega_0^2}\right) - 2\phi\sin\left(\tan^{-1}\frac{\lambda z}{\pi\omega_0^2}\right)\right) \quad (7)$$

As a simplest case example, if the nanoparticle is placed exactly at the focus ($z=0$), the incident field will develop $-\pi/2$ and scattered field will develop $\pi/2$ phase shift in the far-field, hence $\Delta\phi_G(\infty) = \pi$. Then according to Equation (7),

$$E_{out}E_{out}^{*}(z) = E_0^2(1 - 2\frac{\sigma}{A}) \quad (8)$$

This confirms that the coherent addition of the scattered and transmitted beams leads to an effective attenuation of the laser beam in the far-field, as the optical theorem predicts. This also shows the far-field signal is purely from the absorption of the nanoparticle when the nanoparticle is placed at $z = 0$. Also, this argument suggests that in order to obtain an intensity change from the phase shift induced by the nanoparticle, *it has to be displaced from focus*. According to Equation (3), the signal is purely from the phase shift from the nanoparticle when $\Delta\phi_G(\infty) = \frac{\pi}{2}$ or $-\frac{\pi}{2}$. As a whole, this system resembles a Smartt-type interferometer[24] where the two arms follow exactly the same path. Due to this common path nature, the common-mode rejection of various nonideal noise sources such as laser phase noise is improved compared to other types of interferometers. However, this interferometer is inevitably a bright-fringe system, as can be seen from Equation (3), where a small intensity change must be detected on a large DC background.

We now describe how the far-field signal changes depending on the position of the nanoparticle near the focus. The signal-to-noise ratio (SNR) in the shot-noise-limit is proportional to $((\sigma/A)\cos\Delta\phi(z) - 2\phi\sin\Delta\phi(z))\sqrt{P_0/h\nu B}$ with $P_0$ the total laser power at the detector, and $B$ the bandwidth according to Equation (3). To reflect the case where the nanoparticle is displaced from the focus, we can let $\phi = \dfrac{\phi_0}{1+\left(\dfrac{\lambda z}{\pi\omega_0^2}\right)^2}$, $\sigma/A = \dfrac{(\sigma/A)_0}{1+\left(\dfrac{\lambda z}{\pi\omega_0^2}\right)^2}$.

This is because the signals due to the phase shift and the absorption of a single nanoparticle are inversely proportional to the probe beam spotsize area at $z$, and the probe beam area changes with $z$ are assumed to be that of a Gaussian beam. Therefore, the dependence of the SNR on the position $z$ of nanoparticle on axis is proportional to the following:

$$SNR(z) \propto \left|\left(\frac{\sigma}{A}\right)\cos(\Delta\phi(z)) - \phi\sin(\Delta\phi(z))\right| = \frac{\left|\left(\dfrac{\sigma}{A}\right)_0 \cos\left(\tan^{-1}\dfrac{\lambda z}{\pi\omega_0^2}\right) + \phi_0 \sin\left(\tan^{-1}\dfrac{\lambda z}{\pi\omega_0^2}\right)\right|}{1+\left(\dfrac{\lambda z}{\pi\omega_0^2}\right)^2}$$

(9)

This is depicted graphically in Figure 3, where the SNR is represented in arbitrary units. The position of the nanoparticle $z$ is varied from $-2\pi\omega_0^2/\lambda$ to $2\pi\omega_0^2/\lambda$. Assuming for simplicity $\phi + \dfrac{\sigma}{A} = 1$, $\dfrac{\sigma}{A}$ is varied from 0 to 1 to cover a wide range of both absorption and phase shift. This method of plotting is chosen because in the case of a metal nanoparticle or semiconductor quantum dot, the extinction and phase shift at a given wavelength cannot be described accurately by a single transition (where $\phi$ and $\dfrac{\sigma}{A}$ are

connected by Kramers-Kronig relation[25]). Figure 3 (b) presents three special cases:

i: $(\phi, \sigma/A) = (1,0)$, ii: $(\phi, \sigma/A) = (0.5, 0.5)$, iii: $(\phi, \sigma/A) = (0,1)$

**Case i:** $(\phi, \sigma/A) = (1,0)$ This pure phase shift situation is an interesting limit, which can be found in photothermal detection schemes[15] for example. In photothermal detection, the change in refractive index of the sample matrix (or the sample itself for the case of a homogeneous sample) induced by the heat deposited by laser excitation is detected. Therefore, in the case where the heat deposition is modulated in time, the far-field signal detected at the modulation frequency is mainly due to phase shift. Furthermore, if the heat is accumulated around the laser focus by the sample thermal properties, the larger temperature can give an enhanced phase shift. If the thermal diffusivity of the sample matrix is sufficiently low, there can even exist a limit where the signal due to absorption from a nanoparticle is negligible compared to the signal due to phaseshift[26]. In curve i of Fig. 3 (b), the signal is maximum at $z_{max} = \pm 0.71 \pi \omega_0^2 / \lambda$. This result is interesting compared with photothermal lensing spectroscopy[26, 27], where the maximum signal occurs when the homogeneous 2-dimensional sample is placed at $z_{max} = \pm \pi \omega_0^2 / \lambda$. In the case that a diffraction-limited spot is achieved, $z_{max}$ is around 150nm in the visible. This means that in the presence of phase shifting nanoparticles in a very thin layer sample, the far field image will appear to be two parallel planes that are ~300nm apart axially around the focus. If a second laser such as heating laser is used, since the signal is now also proportional to the second laser intensity, $\phi = \dfrac{\phi_0}{\left(1 + \left(\dfrac{\lambda z}{\pi \omega_0^2}\right)^2\right)^2}$ should be used instead of $\phi = \dfrac{\phi_0}{1+\left(\dfrac{\lambda z}{\pi \omega_0^2}\right)^2}$. This assumes for simplicity that the difference between the foci and the beam radii of the two lasers are negligible. In this particular case, the optimal position is given by $z_{max} = \pm 0.5\pi\omega_0^2/\lambda$. This calculation would predict that there would be a discrepancy between the concurrent acquisition of the fluorescence image and the photothermal image since the two measurements will appear to arise from different layers in $z$ of the same sample.

**Case ii:** $(\phi, \sigma/A) = (0.5, 0.5)$ In the case of equal phase and absorption perturbations, curve ii of Fig. 3 (b) shows that the maximum signal is obtained when the particle is placed at positive $z$ ($z = 0.43\pi\omega_0^2/\lambda$). As can be seen in Fig. 3 (a), as the contribution from the absorption signal increases, the peak closer to the laser (negative $z$) decreases in magnitude, while position of the peak closer to the detector approaches the focus. This is because the signal due to extinction interferes destructively with the signal due to phase shift for $z < 0$.

**Case iii:** $(\phi, \sigma/A) = (0,1)$ This is the case when the laser wavelength is on resonance with an absorption line, and no phase shift occurs. Only in this case does the far-field signal behave like one might initially expect: maximized at focus, and decreasing as $z$ moves away from 0.

In conclusion, we have studied the mechanism of the far-field intensity modulation induced by a nanoparticle in a focused Gaussian laser beam. In particular, we have described the exact phase relation of the incident and scattered beam on the optical axis from near-field to far-field. In this analysis, the concept of Gouy phase played a key role in analyzing the behavior of the far-field signal as a function of the nanoparticle position near the focus. One notable result was that a purely dispersive nanoparticle should be displaced from the Gaussian beam focus to generate a far-field intensity change. To comment on the assumptions and consequent limitations of this calculation, first, a transmission geometry in free space was assumed throughout the calculation. The effects of the sample matrix and interfaces with different refractive index are to be further incorporated suiting each experimental situation. Especially, the incident beam can go through another $\pi$ phase shift in a reflection geometry depending on the refractive index contrast. Second, in modern microscopy often a high numerical aperture objective is used, and the focal spot often deviates from Gaussian. In that case, Equation (6) should be modified to an empirical function that saturates faster than tangent hyperbolic.

Considering possible applications, Equation (7) shows that common-path interferometry can be used for measuring the real and imaginary part of the polarizability of a nanoparticle with the capability of obtaining spectra as in Refs. [12][28,29]. If the spectral lineshape of an emitter is well-understood such as in low temperature single-molecule spectroscopy[30], the measurement of the interference signal as in Equation (7) combined with frequency scanning can be used to identify the axial position with an excellent $z$-resolution. Also, it is worth noting that the concept of Gouy phase is robust in the sense that it is valid for any kind of transversely confined beam, such as light from a near-field aperture. Studying the phase aspect in detail of the scattering of a nanoparticle in the optical near-field[29-31][32] would be a worthwhile future effort.

# Figure Captions

**Figure 1.** Schematic of reflection and transmission geometries where the scattering from the nanoparticle interferes with the reflected and transmitted beam, respectively.

**Figure 2.** Schematic of the waist region of a Gaussian beam, showing the position of the waist and the nanoparticle, with the phase accumulations shown at the upper right.

**Figure 3.** (a) Surface plot of the SNR in arbitrary units as a function of normalized nanoparticle position and the fractional absorption, $\sigma/A$. (b) Plot of SNR as a function of normalized nanoparticle position for the three limiting cases described in the text.

# Figure 1

Reflection

Transmission

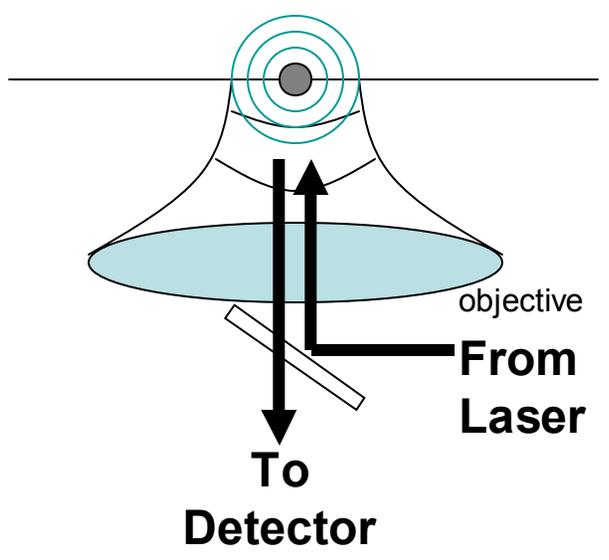
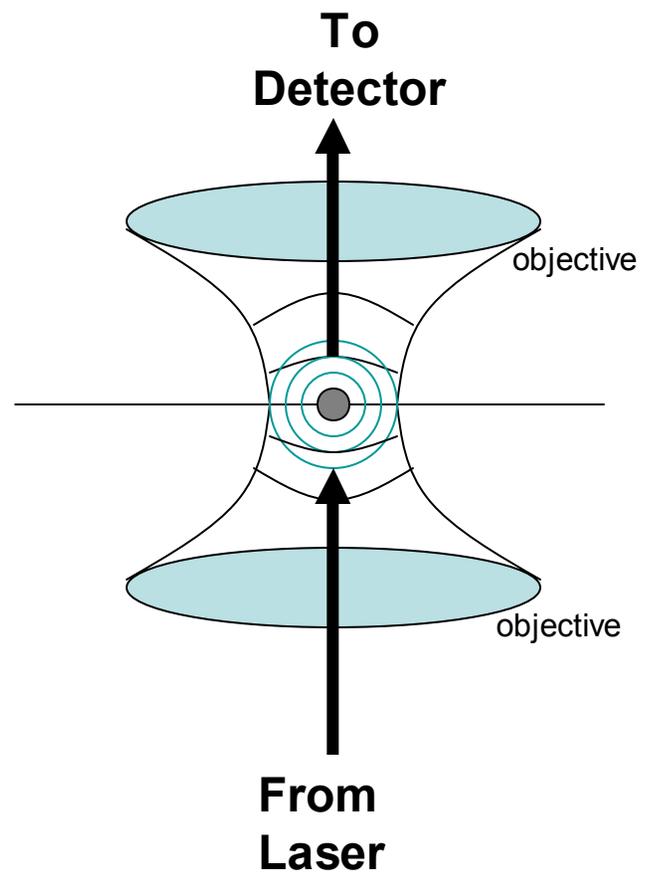

# Figure 2

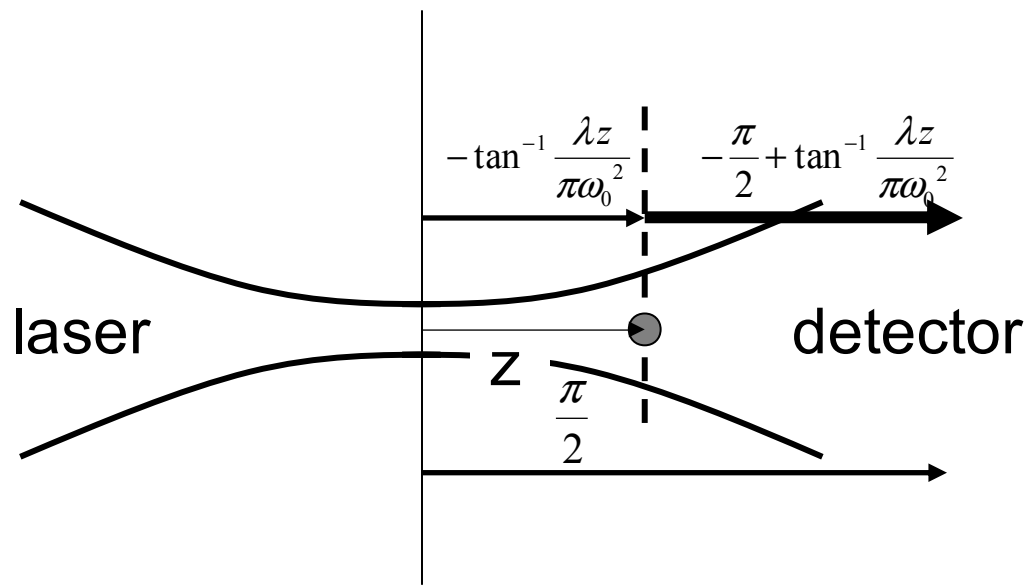

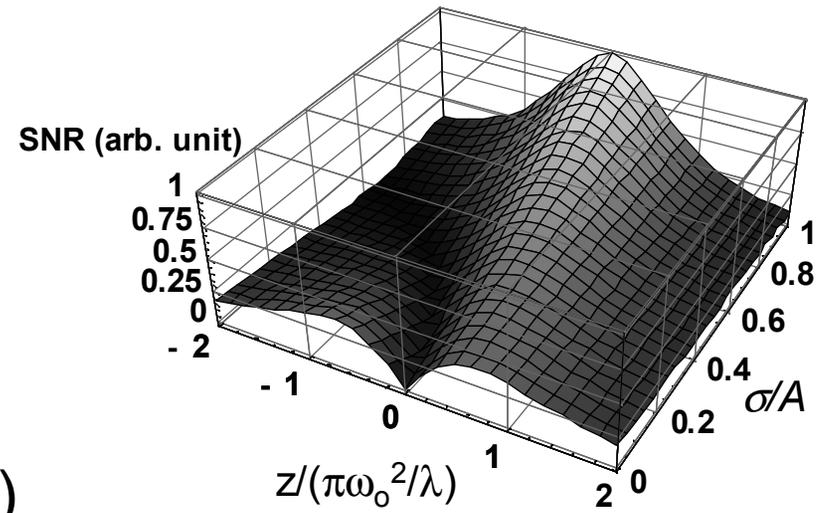
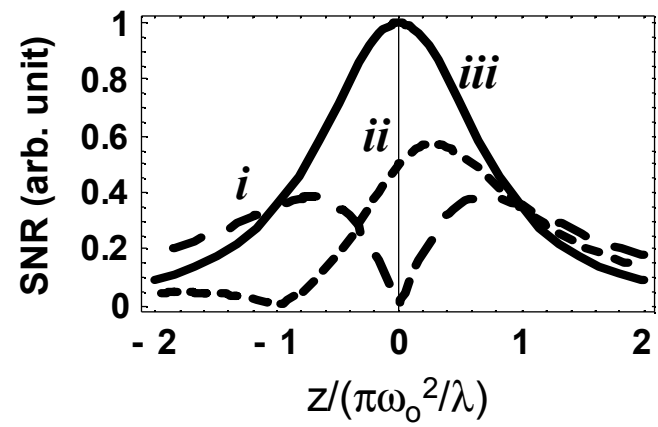